# Interfacial Charge-transfer Excitonic Insulator in a Two-dimensional Organic–inorganic Superlattice

Yang Liu[1,†], Hongen Zhu[2,†], Haifeng Lv[1,†], Yuqiao Guo[1,3,*], Yingcheng Zhao[1], Yue Lin[1], Xiaolin Tai[1], Jiyin Zhao[1], Bingkai Yuan[1,*], Yi Liu[2], Guobin Zhang[2], Zhe Sun[2], Xiaojun Wu[1], Yi Xie[1,3] and Changzheng Wu[1,3,*]

[1] School of Chemistry and Materials Science, CAS Center for Excellence in Nanoscience, CAS Key Laboratory of Mechanical Behavior and Design of Materials, University of Science and Technology of China, Hefei 230026, P. R. China.

[2] National Synchrotron Radiation Laboratory, University of Science and Technology of China, Hefei 230026, P. R. China.

[3] Institute of Energy, Hefei Comprehensive National Science Center, Hefei, Anhui 230031, P. R. China.

[†]*These authors contributed equally: Yang Liu, Hongen Zhu, Haifeng Lv.*

[*]*To whom correspondence should be addressed: E-mail: czwu@ustc.edu.cn, guoyq@ustc.edu.cn, yuanbk@ustc.edu.cn.*

**Excitonic insulators are long-sought-after quantum materials predicted to spontaneously open a gap by the Bose condensation of bound electron-hole pairs, namely, excitons, in their ground state[1-3]. Since the theoretical conjecture, extensive efforts have been devoted to pursuing excitonic insulator platforms for exploring macroscopic quantum phenomena in real materials. Reliable evidences of excitonic character have been obtained in layered chalcogenides as promising candidates[4-9]. However, owing to the interference of intrinsic lattice instabilities[4,8-10], it is still debatable whether those features, such as charge density wave and gap opening, are primarily driven by the excitonic effect or by the lattice transition. Herein, we develop a novel charge-transfer excitonic insulator in organic–inorganic superlattice interfaces, which serves as an ideal platform to decouple the excitonic effect from the lattice effect. In this system, we observe the narrow gap opening and the formation of a charge density wave without periodic lattice distortion, providing visualized evidence of exciton condensation occurring in thermal equilibrium. Our findings identify spontaneous interfacial charge transfer as a new strategy for developing novel excitonic insulators and investigating their correlated many−body physics.**

Excitonic insulators are correlated insulating states arising from the coherent condensation of ground-state excitons and the accompanying gap opening. Excitonic insulators were theoretically predicted to be obtained in semimetals and semiconductors, both with a narrow gap more than sixty years ago[1-3]. Since then, researchers have devoted great efforts to exploring excitonic insulators and exciton condensed phases[4-6,9]. Several layered materials have been identified as the most promising candidates, which exhibit features consistent with excitonic insulators, such as charge density waves (CDWs)[5], anomalous narrow gap openings[7,11] and insulating electrical transport properties[12,13]. In addition to exciton condensation in ground state materials, charge-transfer exciton condensation can be realized in bilayer interfaces comprising two-dimensional (2D) materials under an external stimulus[14-24]. These bilayer systems exhibit excellent controllability of electric, magnetic and light fields, offering powerful methods to engineer exciton condensation and excitonic behaviors[15,20,23-26].

Although continued experimental efforts have been made on excitonic insulators, conclusive experimental evidence for the realization of the ground states of excitonic insulators remains elusive. The main challenge is that exciton condensation induced CDWs in excitonic insulators are nearly impossible to distinguish from period lattice distortion (PLD) induced CDWs because of intrinsic electron-phonon coupling[4,8,9,27]. Therefore, it is difficult to determine whether the insulating gap originates from exciton condensation. To circumvent this problem, exploring new systems with weak electron-phonon coupling[27] and reduced dimensionality[28-30] would provide direct evidence for the realization of excitonic insulators. Organic–inorganic interfaces are a promising platform in which electrons can transfer from organic molecules to inorganic materials and lead to the formation of charge-transfer excitons[31-34]. A recent calculation predicts that high-temperature charge-transfer exciton condensation can be achieved in organic-2D material heterostructures due to the large exciton binding energy[35].

Herein, we report direct evidence that an organic–inorganic superlattice exhibits CDW without PLD and is identified as a new-type charge-transfer excitonic insulator. The superlattice structure is obtained by the self-assembly intercalation of cobaltocene ($Co(Cp)_2$) into the layered chalcogenides $SnSe_2$, in which the separated $SnSe_2$ layers behave as a monolayer. We reveal the presence of a narrow indirect insulating gap of ~90 meV and the existence of a CDW state without PLD, arising from the formation of charge-transfer exciton condensation below the critical temperature of ~225 K. Our results pave the way to discovering excitonic insulators without PLD for exploring quantum phases of ground-state bosons.

We first conceptually describe how ground-state charge-transfer excitons are formed in the $Co(Cp)_2$-$SnSe_2$ superlattice (Fig. 1a,b). The superlattice (Fig. 1c) was prepared by the self-assembly intercalation of $Co(Cp)_2$ molecules based on a solution-based molecular intercalation method (see Methods for details). $Co(Cp)_2$ molecules with 19 valence electrons have intrinsic electron-donating properties; thus, electrons can spontaneously transfer from $Co(Cp)_2$'s highest occupied molecular orbit (HOMO) to the conduction band of $SnSe_2$ (Fig. 1a). The electrons in inorganic $SnSe_2$ layers and the holes in organic Co(Cp)$_2$ layers can pair up to form charge-transfer excitons by Coulomb attraction (Fig. 1b). The charge-transfer excitons are predicted to have large binding energies, long lifetimes and small excitonic masses[35]. These advantages contribute to the formation of exciton condensation below a critical temperature. In addition, the absence of lattice distortion in bulk SnSe$_2$ and the separation of SnSe$_2$ layers by the Co(Cp)$_2$ layer indicate a weak electron-phonon coupling, presenting a new opportunity to decouple the exciton effect from lattice distortion and providing direct evidence for the gap opening mechanism of exciton insulators.

Systematic characterization studies were carried out to investigate the structural information of the as-prepared intercalated sample. We first characterized the structures of pristine SnSe$_2$ crystal and Co(Cp)$_2$-SnSe$_2$ sample using X-ray diffraction (XRD) (Fig. S1). The (001) facet in the SnSe$_2$ crystal shifted significantly toward lower angles after the insertion of Co(Cp)$_2$ into the interlayer, with the interlamellar spacing increasing from 6.1 Å to 11.9 Å. The cross-sectional high−angle annular dark-field scanning transmission electron microscopy (HAADF-STEM) image further confirmed that the interlayer spacing increased to 11.9 Å in the intercalated sample (Fig. 1c), which agrees well with the XRD studies. The increased interlayer spacing (5.8 Å) is consistent with the molecular size, implying the successful intercalation of Co(Cp)$_2$ into SnSe$_2$ layers. Since the adjacent SnSe$_2$ layers are isolated by the intercalated Co(Cp)$_2$, Raman modes of SnSe$_2$ in the Co(Cp)$_2$−SnSe$_2$ sample are consistent with the values in monolayer SnSe$_2$ (Fig. S2)[36], suggesting a weak interlayer coupling.

Soft X-ray absorption spectroscopy (sXAS) studies were employed to identify the existence of spontaneous interfacial charge transfer in the superlattice (Fig. S3). After Co(Cp)$_2$ intercalation, Co-L edge spectra present a shift of absorption peaks to higher energy values, revealing a decrease in valence electrons and electron transfer from Co(Cp)$_2$ to SnSe$_2$. The electron transfer from Co(Cp)$_2$ to SnSe$_2$ is also supported by the theoretical charge analysis, which shows charge transfer from Co(Cp)$_2$ to SnSe$_2$ with a value of 0.40 e$^-$ per unit cell.

Having identified the interlayer structure, we further studied the intralyer structures of Co(Cp)$_2$-SnSe$_2$ superlattice. Electron probe microanalysis (EPMA) was performed to determine the compositional distribution. Elemental mapping for Sn, Se and Co revealed a spatially homogeneous distribution (Fig. S4). Quantitative analysis revealed that the Co/Sn ratio was ~0.313 in the intercalated sample (Table S1), which was in good agreement with the previous theoretical prediction of $SnSe_2\{Co(Cp)_2\}_{1/3}$ superlattice[33]. The assembly structure of $Co(Cp)_2$ on the atomic scale was characterized using scanning tunneling microscopy (STM). The $Co(Cp)_2$-$SnSe_2$ superlattice is prepared by cleaving the sample in ultrahigh vacuum prior to measurements (Fig. 1e). The STM topographies show that the cleaved surface is a long-range ordered $Co(Cp)_2$ layer (Fig. 1f). In the STM image, a single $Co(Cp)_2$ molecule appears as ellipse-like protrusions and adopts a lying configuration. This is consistent with previous studies based on various techniques[37] and theoretical calculations[33] that all the intercalated $Co(Cp)_2$ molecules adopt a preferential configuration with the metal-to-ring axes parallel to the $SnSe_2$ planes. It is noted that structural defects such as $Co(Cp)_2$ molecules in different adsorption directions are observed in the $Co(Cp)_2$ layer, as marked by gray dotted ovals in Fig. 1f.

In summary, our structural characterization shows that the $Co(Cp)_2$ molecules are intercalated between every $SnSe_2$ layer, forming a $Co(Cp)_2$-$SnSe_2$ superlattice (Fig. 1c). The intercalated $Co(Cp)_2$ molecules form long-range ordered layers with a molecular axis parallel to the $SnSe_2$ layer. The separated $SnSe_2$ layers behave as a monolayer and indicate a reduced dimensionality, which is beneficial to the suppression of the screening effect[36].

One of the main features of excitonic insulators is a CDW state in real space induced by excitonic instability without PLD[4,27]. Using STM and non-contact atomic force microscopy (NC-AFM), we determined the emergence of CDW state in the $Co(Cp)_2$-$SnSe_2$ superlattice. For pristine $SnSe_2$, a hexagonal lattice is observed in atomic resolved STM and NC-AFM images (Fig. 2a and Fig. S6), corresponding to the top layer of Se atoms. After $Co(Cp)_2$ intercalation, a complex CDW emerges on $SnSe_2$ at negative bias (Fig. 2b,c), while the image at +0.5 V shows only a hexagonal lattice (Fig. S5). The CDW at negative bias existed with the atomic lattice and had a periodic structure with a monoclinic unit cell (1.568 nm/1.265 nm, degree: 65.9°). The STM images show that Se atom rows have alternate bright and dim contrast in zigzag rows, as marked in Fig 2b. In addition, Se atoms within a unit cell exhibit rich fine structures and show different brightness levels, implying non-uniform electronic modulation from Co(Cp)$_2$. It is found that non-uniform electronic modulation also exists in a large range, such as those marked by red cycles in Fig 2c. Considering the incommensurate structure between Co(Cp)$_2$ and SnSe$_2$ and the defects in the Co(Cp)$_2$ layer, the rich fine structure and non-uniform electronic distribution within the CDW state may result from these structural imperfections.

To identify an intrinsic excitonic insulator, it is necessary to exclude the influence of PLD. In general, CDW is accompanied by PLD through the Peierls transition[4,5,10,27]. Thus, it is difficult to distinguish between PLD induced CDW and exciton condensation induced CDW. To discriminate the mechanism of CDW, we further performed NC-AFM measurements to determine the lattice structure on the atomic scale. Fig. 2d is the constant-height frequency shift (Δf) NC-AFM image of SnSe$_2$ acquired in the Pauli repulsion area, which is the same area as that in Fig. 2b and c. The AFM image shows a hexagonal lattice without PLD and the positions of dark holes (maximum attraction) correspond to the top Se atoms in the SnSe$_2$ layer. Generally, STM images reflect both electronic states near the Fermi level and geometries of the surface. The contrast in NC-AFM images mainly arises from Pauli repulsion at close tip-sample distances, which increases with total electron density and reflects the geometric position of atoms[38-40]. Hence, considering the CDW observed in the STM images and the absence of PLD in the AFM image, we suggest that the observed CDW originates from charge-transfer exciton condensation. This result means that Co(Cp)$_2$-SnSe$_2$ superlattice have weak electron-phonon coupling, offering an ideal platform for decoupling the exciton effect from lattice distortion.

To examine the electronic structures before and after intercalation, we performed angle-resolved photoemission spectroscopy (ARPES) measurements on the pristine SnSe$_2$ crystal and Co(Cp)$_2$-SnSe$_2$ superlattice. The ARPES spectra were measured in the energy-momentum space along the Γ-M direction at 20 K (Fig. 3b). Different from the semiconductor behavior of bulk SnSe$_2$ with a large gap (Fig. 3c, matched with the calculated band structure of Fig. S7a), the band structure of $Co(Cp)_2$-$SnSe_2$ superlattice shows that an electron pocket appears in the Fermi surface around the M point, and the conduction band at the M point shifts downwards and crosses Fermi level ($E_F$) (Fig. 3d and Fig. S8), indicating charge transfer from the HOMO of $Co(Cp)_2$ to the conduction band of $SnSe_2$ (Fig. 3a). Meanwhile, as shown in Fig. 3e, an extra valence band is introduced into the original gap of host $SnSe_2$, which belongs to $Co(Cp)_2$ molecules. This in-gap band is consistent with the band structure obtained by theoretical calculations (Fig. S7c). In addition, the states below -2 eV (Fig. 3d) conform to the bands of monolayer $SnSe_2$ (Fig. S7b), which further confirms that each $SnSe_2$ layer behaves as a single layer and exhibits a reduced dimensionality, contributing to exciton condensation[36].

Another main feature of excitonic insulators is the insulating gap induced by exciton condensation below a critical temperature[1-3,5,10,11]. Here, the insulating gap ($E_g$) means the energy difference between the electron's conduction band minimum (CBM) and the hole's valence band maximum (VBM). The detailed analysis of the electron pocket around the M point indeed shows the presence of an insulating gap opening (Fig. 3e). CBM and VBM are located at slightly different positions, $k_x = \sim 1.023$ 1/Å and $k_x = \sim 1.098$ 1/Å, respectively, indicating that the excitonic gap is an indirect band gap separated by a wave vector $q_0$. The value of $E_g$ can be estimated from the analysis of the ARPES intensity in terms of energy distribution curves (EDCs) at the CBM and VBM (Fig. 3f). $E_g$ is determined to be ~90 meV, which is a narrow gap favorable for the formation of excitonic insulators. Furthermore, the wave vector in reciprocal space corresponds to a CDW period of L = ~8.4 nm ($L=2\pi/q_0$)[4]. It is noted that the period deduced from the wave vector shows a deviation from that in STM images. This deviation may originate from the incommensurate structure between $Co(Cp)_2$ and $SnSe_2$ and the defects in the $Co(Cp)_2$ layer, and these structural imperfections would result in a larger CDW period.

The metal-insulator transition in electrical transport behaviors is a physical observable of excitonic insulators[2]. We measured the temperature-dependent resistivity of the pristine $SnSe_2$ crystal and $Co(Cp)_2$-$SnSe_2$ superlattice using a physical property measurement system (PPMS). Different from pristine SnSe$_2$ with semiconductor behavior, Co(Cp)$_2$-SnSe$_2$ superlattice exhibits abnormal electrical transport behaviors. With decreasing temperature, the superlattice successively experiences metallic, anomalous insulating and superconducting states (Fig. 4a)[33,41]. From room temperature to 225 K, the resistance decreases with temperature, showing metallic behavior. A resistive anomaly appears at ~225 K, showing a metal-insulator transition and the presence of CDW. This is consistent with the observed CDW by STM, suggesting that charge-transfer excitons begin to condense and result in the insulating state. The critical temperature of exciton condensation ($T_d$, ~225 K) is related to the exciton binding energy ($E_b$): $kT_d \approx 0.1 E_b$[42]. The calculated $E_b$ is approximately 190 meV, meeting the crucial criterion of $E_b > E_g$ in excitonic insulators.

Unexpectedly, we observed the emergence of superconductivity at a $T_c$ of 4.8 K in the superlattice, indicating a transition between two coherent bosonic states: exciton condensation and superconductivity, as marked by the red dashed line in Fig. 4b. The superconductivity was further demonstrated by the Meissner effect in the low-temperature field-dependent magnetization (M-H) curves (Fig. S9). It was theoretically predicted that the model of excitonic insulators could exhibit the electrical properties of a superconductor[43]. A similar transition behavior was reported in gate-controlled TiSe$_2$[44] or pressure-modulated Ta$_2$NiSe$_5$[45], whereas such a transition at ground state without external modulations was first observed. The in-depth understanding of the transition needs to be explored by future research.

In summary, we have developed a novel charge-transfer excitonic insulator in an organic–inorganic superlattice. By combining STM, AFM and ARPES measurements, we have experimentally observed CDW without PLD and a small excitonic gap, providing direct evidence of exciton condensation. Furthermore, excitonic insulators can correlate with many exotic phenomena, such as superconductors[43], topological insulators[29,30,46] and nematic insulators[47]. In our study, an impressive phase transition between the two bosonic states is observed, providing an ideal platform for exploring the collective behaviors emerging from multiple composite bosons.

**Acknowledgement**

This work was financially supported by the National Natural Science Foundation of China (U2032161, 21925110, 21890750, 12147105), USTC Research Funds of the Double First-Class Initiative (YD2060002004), the Youth Innovation Promotion Association CAS (2018500), the Strategic Priority Research Program of Chinese Academy of Sciences (XDB36000000), Key R&D Program of Shandong Province (2021CXGC010302) and the Users with Excellence Project of Hefei Science Center CAS (2021HSC-UE004). We appreciate the support from beamline 1W1B of the Beijing Synchrotron Radiation Facility (BSRF, Beijing, China) and beamlines BL13U and BL12B-a of the National Synchrotron Radiation Laboratory (NSRL, Hefei, China). We thank the Cryo-EM Center at the University of Science and Technology of China for the EM facility support and the support from the Super Computer Centre of USTCSCC and SCCAS. This work was partially carried out at the USTC Center for Micro and Nanoscale Research and Fabrication.

**Contributions**

C.W. supervised the entire project and is responsible for the infrastructure and project direction, conceived the idea, co-experimentally realized the study and co-wrote the paper, Y.G. conceived the idea, co-experimentally realized the study, co-wrote the paper and supervised the entire project. Y.L., H.Z. and H.L. contributed equally to this work; they experimentally realized the study, analysed the data and co-wrote the paper. These works were assisted by Y.L., X.T., J. Z, B.Y., Y.L., G.Z., Z.S., X.W. STM and AFM data collection and analysis were performed by B.Y. HAADF-STEM data collection was performed by Y.L. and X.T. ARPES data collection and analysis were performed by H.Z., Y.L., G.Z. and Z.S. Theoretical calculations were carried out by H.L. and X.W. Y.X. supervised the whole experimental procedure and co-wrote the paper. All authors discussed the results and commented on and revised the manuscript.

**Competing interests**

The authors declare no competing financial interests.

**Figures**

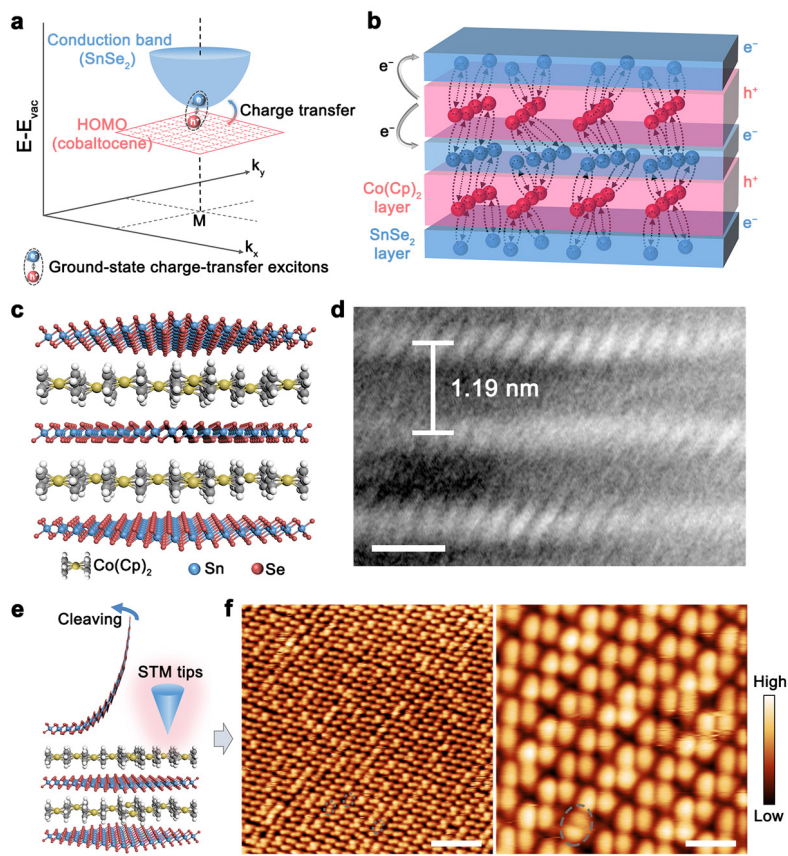

**Fig. 1 | Interfacial charge-transfer excitonic insulator with an organic–inorganic superlattice. a**, **b**, Schematic drawings of the forming mechanism of ground-state interfacial excitons from reciprocal space and real space perspectives, respectively. **c**, Model diagram of the $Co(Cp)_2$-$SnSe_2$ superlattice structure. **d**, Cross-sectional HAADF-STEM image of the $Co(Cp)_2$-$SnSe_2$ superlattice. Scale bar, 1 nm. **e**, Schematic diagram of cleaving the $Co(Cp)_2$-$SnSe_2$ superlattice along the $SnSe_2$ layer to expose the $Co(Cp)_2$ layer. **f**, STM images of the exposed $Co(Cp)_2$ layer acquired at 78 K. Scale bars in f (the left and the right): 8 nm and 2 nm, respectively.

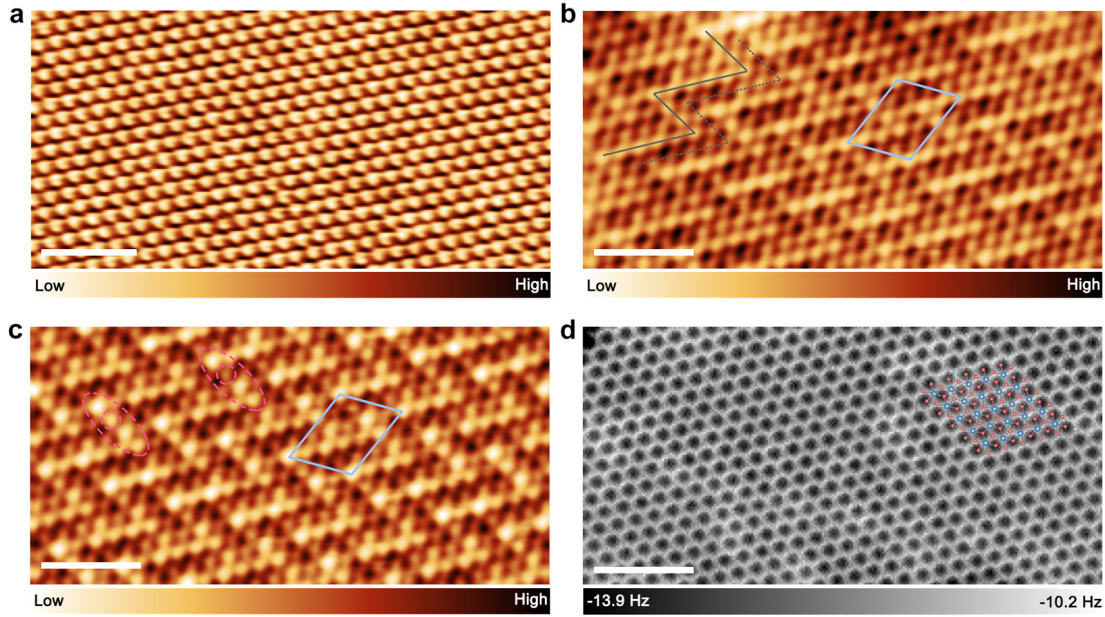

**Fig. 2 | STM and NC-AFM characterization of the electronic and structural properties of the SnSe$_2$ layers in pristine SnSe$_2$ crystal and the Co(Cp)$_2$-SnSe$_2$ superlattice. a**, Constant-current STM images of pristine SnSe$_2$ (V = -1.0 V, I = 5 pA). **b**, **c**, Constant-current STM images of the SnSe$_2$ layer in the Co(Cp)$_2$-SnSe$_2$ superlattice acquired at different biases (**b**: V = -1.0 V, I = 5 pA; **c**: V = -1.5 V, I = 2 pA). **d**, Constant-height AFM (Δf) image of SnSe$_2$ of the same area as **b** and **c** at tip height of +100 pm. The tip height is referenced to the STM setpoint (V = 0.5 V, I = 5 pA) on the SnSe$_2$ surface. The structure of SnSe$_2$ is overlaid on the AFM image, in which red and blue dots represent Se and Sn atoms, respectively. All the data are acquired at 4.8 K. Scale bars: 2 nm.

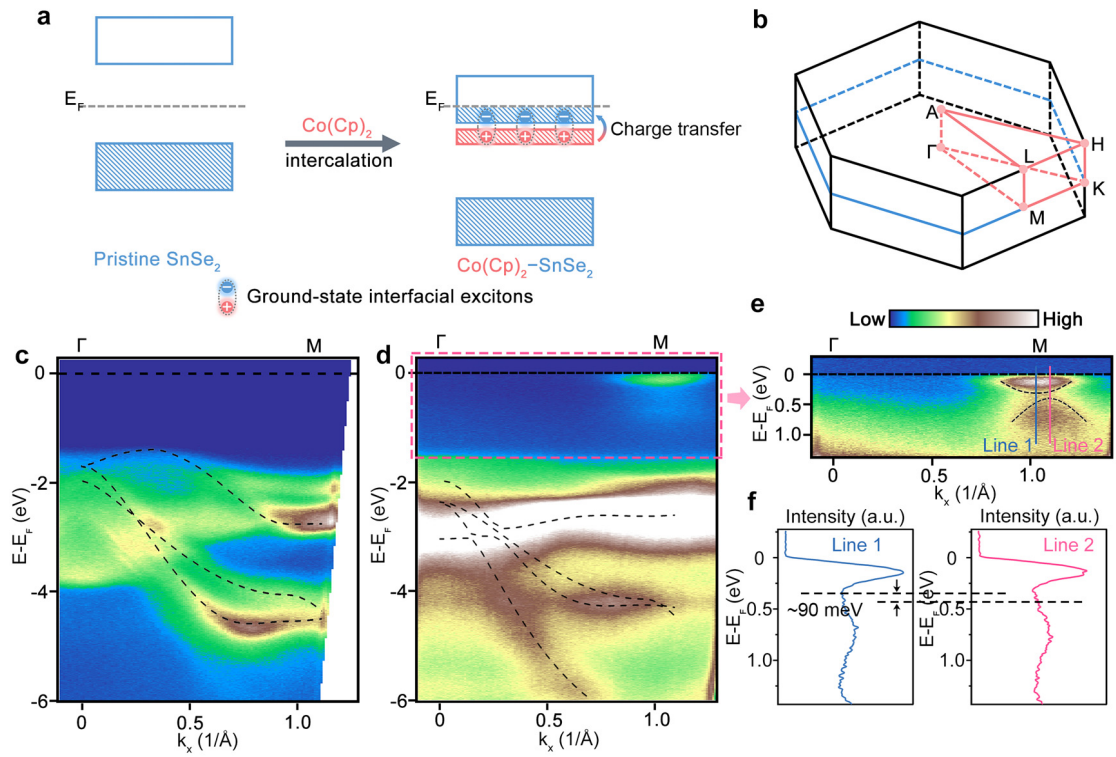

**Fig. 3 | Electronic band structures of pristine SnSe$_2$ crystal and Co(Cp)$_2$-SnSe$_2$ superlattice. a**, Schematic diagram of charge-transfer exciton formation from the perspective of band structure. **b**, The hexagonal Brillouin zone and high-symmetry points. **c**, **d**, ARPES intensity plots of pristine SnSe$_2$ and Co(Cp)$_2$-SnSe$_2$ superlattice, respectively. **e**, Enlarged view of the red dashed box in **d** to identify the excitonic gap. **f**, EDCs taken along Line 1 and Line 2 in **e**.

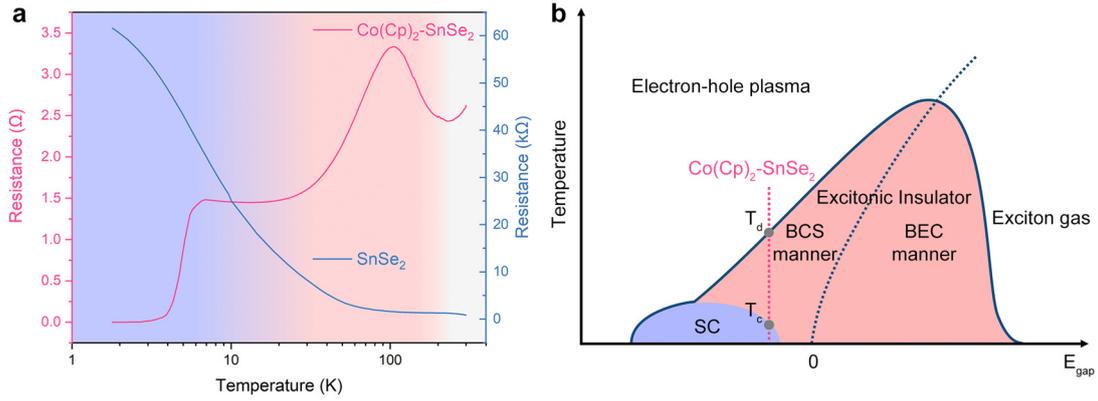

**Fig. 4 | Phase transition of Co(Cp)$_2$-SnSe$_2$ superlattice. a**, electrical transport properties of SnSe$_2$ crystal and Co(Cp)$_2$-SnSe$_2$ superlattice. **b**, Phase transitions of Co(Cp)$_2$- SnSe$_2$ superlattice in phase diagram of excitonic insulator as indicated by red dotted line. In **b**, BCS, BEC and SC are short for Bardeen-Cooper-Schrieffer, Bose–Einstein condensation and superconductor, respectively.